\documentclass[twocolumn,showpacs,preprintnumbers,amsmath,amssymb]{revtex4}

\usepackage{graphicx}
\usepackage{dcolumn}
\usepackage{bm}

\def\bra#1{\mathinner{\langle{#1}|}}
\def\ket#1{\mathinner{|{#1}\rangle}}
\def\rbra#1{\mathinner{\langle{#1}\parallel}}
\def\rket#1{\mathinner{\parallel{#1}\rangle}}

\def\Bra#1{\left<#1\right|}
\def\Ket#1{\left|#1\right>}
\newcommand{\Heff}{\ensuremath{\mathrm{H}_{\mbox{\tiny eff}}}}
\newcommand{\Hhf}{\ensuremath{\mathrm{H}_{\mbox{\tiny hf}}}}
\newcommand{\HSt}{\ensuremath{\mathrm{H}_{\mbox{\tiny St}}}}

\newcommand{\Hz}{\mbox{Hz}}

\newcommand{\cm}{\mbox{cm}}
\newcommand{\kV}{\mbox{kV}}

\begin{document}

\preprint{APS/123-QED}

\title{Stark effect of the cesium ground state:\\ electric tensor
polarizability and shift of the clock transition frequency}
\author{S. Ulzega}
\email{simone.ulzega@unifr.ch}
\author{A. Hofer}
\author{P. Moroshkin}
\author{A. Weis}
\affiliation{%
Physics Department, Fribourg University, Chemin du Mus\'ee 3, 1700
Fribourg, Switzerland
}%

\date{\today}

\begin{abstract}

We present a theoretical analysis of the Stark effect in the
hyperfine structure of the cesium ground-state.
We have used third order perturbation theory, including diagonal and
off-diagonal hyperfine interactions, and have identified terms which
were not considered in earlier treatments.
A numerical evaluation using perturbing levels up to $n=18$ yields new
values for the tensor polarizability $\alpha_2(6S_{1/2})$ and for the
Stark shift of the clock transition frequency in cesium.
The polarizabilities are in good agreement with experimental values,
thereby removing a 40-year-old discrepancy.
%
The clock shift value is in excellent agreement with a recent
measurement, but in contradiction with the commonly accepted value
used to correct the black-body shift of primary frequency standards.

\end{abstract}

\maketitle

Since its discovery, the Stark effect has been an important
spectroscopic tool for elucidating atomic structure.
The Zeeman degeneracy of the $nS_{1/2}$ ground state of alkali
atoms cannot be lifted by a static electric field because of time
reversal invariance.
However, the joint effect of the hyperfine interaction and the
Stark interaction leads to both $F$ and $\left|M\right|$ dependent
energy shifts which in cesium are 5~and 7~orders of magnitude
smaller than the shift due to the second order scalar
polarizability.
%
%
While the scalar Stark shift is understood at a level of $\left( 1
\mbox{--} 2 \right)\cdot
10^{-3}$~\cite{amini_scalpol,derevianko_scalpol,zhou_scalpol},
there has so far been no satisfactory theoretical description of
the $F$ and $\left|M\right|$ dependent alterations of the scalar
effect at the levels of $10^{-5}$ and $10^{-7}$.
In this Letter, we extend previous theoretical models which allows
us to bridge a long standing gap between theory and experiment.

The effect of a static electric field on the hyperfine structure
was treated in a comprehensive paper by Angel and
Sandars~\cite{angel_sandars}, who showed that the second order
Stark effect of a state $\ket{\gamma}=\ket{nL_J,\,F,M}$ can be
parametrized in terms of a scalar polarizability,
$\alpha_0^{(2)}(\gamma)$, and a tensor polarizability,
$\alpha_2^{(2)}(\gamma)$, where the latter has non-zero values for
states with $L~\geq~1$ only.
As a consequence, the spherically symmetric $nS_{1/2}$ alkali-atom
ground state has only a scalar polarizability, so that its magnetic
sublevels $\ket{F,M}$ all experience the same Stark shift,
independent of $F$ and $M$.
However, an experiment by Haun and Zacharias in
1957~\cite{haun_zacharias} showed that a static electric field
induces a quadratic (in field strength $\mathbb{E}$) shift of the
$\ket{F=4,M=0} \leftrightarrow \ket{F=3,M=0}$ hyperfine (clock)
transition frequency ($F$-dependent effect).
In 1964, Lipworth and Sandars~\cite{lipworth_sandars} demonstrated
that a static electric field also lifts the Zeeman degeneracy within
the $F=4$ sublevel manifold of the cesium ground state ($M$-dependent
effect).
An improved measurement was performed later by Carrico
et~al.~\cite{carrico_tenspol} and its extension to five stable alkali
isotopes was done by Gould et~al.~\cite{gould_tenspol}.

Sandars~\cite{sandars_hatom} has shown that the $F$- and
$M$-dependence of the Stark effect can be explained when the
perturbation theory is extended to third order after including the
hyperfine interaction.
The theoretical expressions for the tensor polarizabilties
$\alpha_{2}$ of~\cite{sandars_hatom} were evaluated numerically
in~\cite{lipworth_sandars} and~\cite{gould_tenspol} under
simplifying assumptions.
Comparison with the experimental polarizabilities showed that the
absolute theoretical values were systematically larger for all
five alkalis studied in~\cite{gould_tenspol}.
Our recent
measurements~\cite{ospelkaus_tenspol,ulzega,UlzegaPhDthesis} of
$\alpha_{2}$ of $^{133}\text{Cs}$ confirmed the earlier
experimental values~\cite{carrico_tenspol,gould_tenspol}.
To help clarify this long-standing discrepancy we have reanalyzed
the third-order calculation of the Stark effect in the alkali
hyperfine structure.
We have identified contributions which were not included in
earlier calculations and as a result we obtain theoretical values
which are in good agreement with experimental
results~\cite{carrico_tenspol,gould_tenspol,ospelkaus_tenspol,ulzega,UlzegaPhDthesis}.
When applied to the Stark shift of the clock transition frequency
our calculations yield a value in good agreement with a recent
measurement~\cite{godone}.

The third order perturbation of the energy $E_{\alpha}$ of the
sublevel $\ket{\alpha} = \ket{6S_{1/2},F,M}$
is given by
\begin{multline}
\Delta E^{(3)}(\alpha)
=\sum_{\beta\neq\alpha, \gamma\neq\alpha}
\frac{\Bra{\alpha} W \Ket{ \beta}
      \Bra{ \beta} W \Ket{\gamma}
      \Bra{\gamma} W \Ket{\alpha}}
     {(E_{\alpha}-E_{\beta})(E_{\alpha}-E_{\gamma})}\\
- \Bra{\alpha}W \Ket{\alpha} \sum_{\beta \neq \alpha}
\frac{\left| \Bra{\beta}W\Ket{\alpha} \right|^{2}}{(E_{\alpha}-E_{\beta})^2}
\:,
\label{thirdordergeneral}
\end{multline}
where $E_{\beta}$ and $E_{\gamma}$ are the energies of unperturbed
states, and the perturbation $W=\Hhf+\HSt$ describes the hyperfine
and Stark interactions.
Of all the terms obtained by substituting W into
Eq.~(\ref{thirdordergeneral}) only those containing the product of
two matrix elements of \HSt{} and one matrix element of \Hhf{}
give nonzero contributions because of the selection rules $\Delta
L=\pm 1$ for \HSt{} and $\Delta L=0,\pm 2$ for \Hhf{}.

\begin{figure}[t]
\includegraphics[width=8cm]{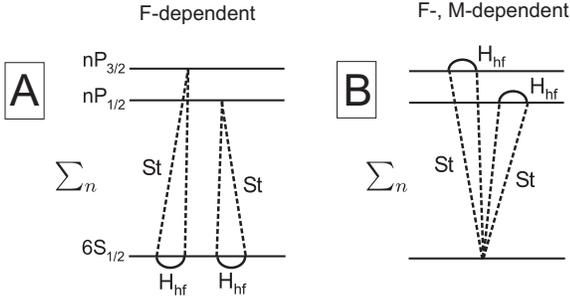}
\caption{Contribution of diagonal hyperfine matrix elements to the
third order Stark effect calculation.  The dotted lines
represent Stark interaction matrix elements while the solid
lines represent hyperfine matrix elements.}
\label{fig:diagmatel}
\end{figure}

We address first the second term of Eq.~(\ref{thirdordergeneral}).
The diagonal matrix element in front of the sum represents only
the Fermi contact interaction in the ground state, while the sum
is carried out over Stark interactions only.
Diagram $\textbf{A}$ of Fig.~\ref{fig:diagmatel} represents this term
in graphical form.
The hyperfine matrix element $\bra{\alpha} \Hhf \ket{\alpha}$ is
scalar, but $F$ dependent, and the sum is similar to the
expression for the second order polarizability, except for the
squared energy denominator.
As in the theory of the second order
polarizability~\cite{angel_sandars} the sum can therefore be
expressed as $\Bra{\alpha} \Heff^{(k=0)}+\Heff^{(k=2)} \Ket{\alpha}$.
The scalar contribution, $\bra{\alpha} \Heff^{(0)} \ket{\alpha}$,
depends only on the strength of the applied electric field,
$\mathbb{E}^{2}$, while the second rank tensor contribution
$\Heff^{(2)}$ depends on its orientation as
$3\mathbb{E}_{z}^{2}-\mathbb{E}^{2}$.
The selection rules for tensor operators require that
$\rbra{S_{1/2}} \Heff^{(2)} \rket{S_{1/2}}=0$.
As a consequence the second term of Eq.~(\ref{thirdordergeneral})
gives a scalar contribution to the energy which depends on $F$,
but not on~$M$, and which can be parametrized by a~third order
scalar polarizability $\alpha_{0}^{(3)}(6S_{1/2},F)$ as
\begin{equation}
\Delta E^{(3)}_0(F)=-\frac{1}{2} \cdot \alpha_{0}^{(3)}(F) \cdot \mathbb{E}^{2} \,.
\label{scalpol3}
\end{equation}
This term produces the major contribution to the Stark shift of
the hyperfine transition frequency
$\mathinner\Delta\nu_{00}\!=\!\nu(F\!=\!4,
M\!=\!0)-\nu(F\!=\!3,M\!=\!0)$.
\begin{figure}[t]
\includegraphics[width=8cm]{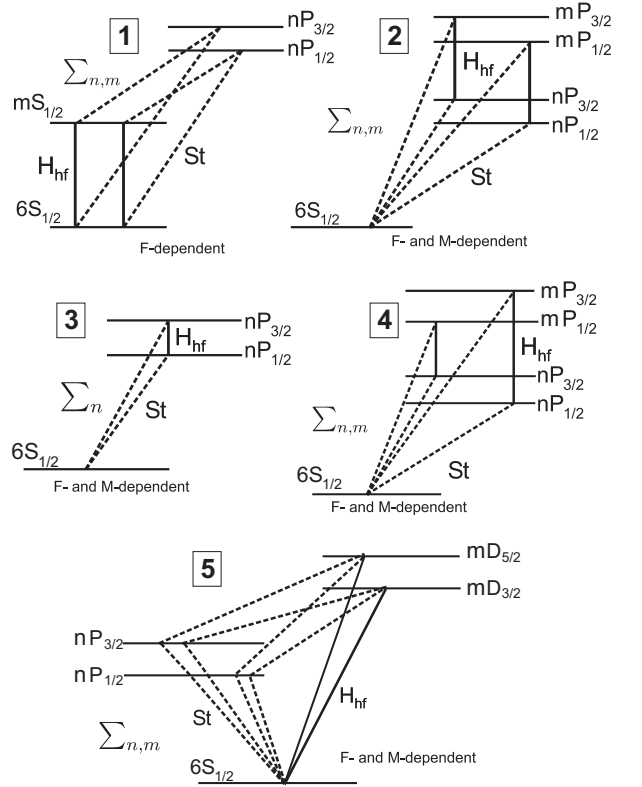}
\caption{Contribution of off-diagonal hyperfine matrix elements to
the third order calculation of the Stark effect.}
\label{fig:offdiagmatel}
\end{figure}

We next address the first term of Eq.~(\ref{thirdordergeneral})
and consider only diagonal matrix elements of \Hhf.
As before, the Stark interaction part of the first term
has only a rank $k=0$ (scalar) contribution.
The dipole-dipole and electric quadrupole parts of the hyperfine
interaction have the rotational symmetry of $k=0,2$ and $k=2$
tensors, respectively.
Together with the scalar Stark interaction, the first term thus has
scalar and tensor parts.
The scalar part has the same $F$ dependence~\cite{UlzegaPhDthesis} as
$\alpha_{0}^{(3)}(F)$, and corrects the latter by approximately 1\%,
while the second rank tensor parts have an $F$ and $M$ dependence,
which can be parametrized in terms of a third order tensor
polarizability $\alpha_{2}^{(3)}(6S_{1/2},F)$ via
\begin{multline}
\Delta E_{2}^{(3)}(\alpha)= -\frac{1}{2}
\alpha_{2}^{(3)}(F)\frac{3M^{2}-F(F+1)}{2 I(2I+1)}
f(\theta)\mathbb{E}^{2} \,, \label{tensorpolarizability}
\end{multline}
where the dependence on the angle $\theta$ between the electric
field and the quantization axis is given by
$f(\theta)=3\cos^{2}\theta-1$.
The $M$-dependence in Eq.~(\ref{tensorpolarizability}) is
responsible for lifting the Zeeman degeneracy in the ground state
hyperfine levels, but gives also a correction to the shift of the
clock transition frequency which itself is dominated by
Eq.~(\ref{scalpol3}). The third order Stark effect of the two
hyperfine levels can then be parametrized in terms of the third
order polarizabilities by
\begin{multline}
\alpha^{(3)}(F,M)=\alpha_{0}^{(3)}(F) + \alpha_{2}^{(3)}(F)
\frac{3M^{2}\!-\!F(F\!+\!1)}{2 I(2I+1)} f(\theta) \,.
\label{globaleffect}
\end{multline}
In cesium the explicit $F$-dependence of Eq.~(\ref{globaleffect})
yields~\cite{ulzega,UlzegaPhDthesis}, for $\theta=0$,
\begin{subequations}
\label{eq:alpha2Sandras}
\begin{eqnarray}
\alpha^{(3)}(4,M)\!&=&\!a _0+\left(a_1+a_2\right)
\frac{3M^{2}\!-\!20}{28}\:,
\label{subeq:alpha2F=4}\\
\alpha^{(3)}(3,M)\!&=&\!-\frac{9}{7}a_0+\left(-a_1+
\frac{5}{3}a_2\right) \frac{3M^{2}\!-\!12}{28}\:,
\label{subeq:alpha2F=3}
\end{eqnarray}
\end{subequations}
where $a_1$ is the contribution of the tensor part of the
dipole-dipole hyperfine interaction, and $a_2$ the contribution of
the electric quadrupole interaction.
The Fermi-contact interaction provides the dominant contribution
to $a_0$ which also has a~small contribution from the scalar part
of the dipole-dipole interaction.
Equations~(\ref{eq:alpha2Sandras}) bear a close resemblance to the
expressions obtained by Sandars~\cite{sandars_hatom}, except for
the negative sign of the $a_1$ term in
Eq.~(\ref{subeq:alpha2F=3}), which is positive in Sandars work.
We have confirmed the sign of our expression in a recent
experiment~\cite{ulzega}.
That sign error, which seems to have remained unnoticed for almost 40
years, has no consequence for the tensor polarizability of the $F=4$
state, which is the only one measured to date.
It affects the static Stark shift of the clock transition at a
level slightly below today's experimental sensitivity.

\begin{figure}[b]
\includegraphics[width=7cm]{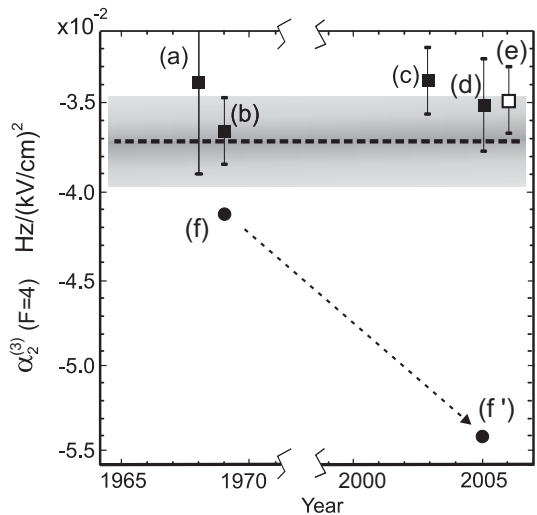}
\caption{The third order tensor polarizability of the F=4 Cs
ground state. The filled squares ({\tiny$\blacksquare$}) represent
experimental values of Carrico et~al.~\cite{carrico_tenspol}(a),
Gould et~al.~\cite{gould_tenspol}(b), Ospelkaus
et~al.~\cite{ospelkaus_tenspol}(c) and Ulzega
et~al.~\cite{ulzega}(d). The empty square ({\tiny$\square$}) (e)
represents a weighted average of (a), (b), (c) and (d). The dots
(\textbullet) represent the theoretical value
from~\cite{gould_tenspol}(f), and our re-evaluation of the latter
value after dropping simplifying assumptions (f').  The dotted
horizontal line is the result of the present work with its
uncertainty (shaded band).} \label{fig:tenspolstory}
\end{figure}

Sandars' equations were evaluated by~\cite{lipworth_sandars}
and~\cite{gould_tenspol} who considered only \emph{diagonal}
matrix elements of \Hhf{} for the $6S_{1/2}$ and the $6P_J$
states (Fig.~\ref{fig:diagmatel}).
They further neglected the fine structure energy splitting in the
denominators of Eq.~(\ref{thirdordergeneral}) and assumed the
relation $A_{6P_{1/2}} = 5 A_{6P_{3/2}}$ for the hyperfine
constants, valid for one-electron atoms, while for Cs the
corresponding ratio of experimental values is~5.8.
Those assumptions yielded the value (f) in
Fig.~\ref{fig:tenspolstory}, which is in disagreement with the
experimental results.
We have reevaluated~\cite{UlzegaPhDthesis} their result by
dropping the simplifying assumptions and by using recent
experimental values of the reduced matrix elements
$\rbra{6S_{1/2}}d \rket{6P_{j}}$
from~\cite{rafac_tanner_redmatel}.
As a result, the discrepancy becomes even larger [point (f') in
Fig.~\ref{fig:tenspolstory}], and does not change significantly
when the perturbation sum is extended to $nP_J$ states with $n>6$.

\begin{figure}[t]
\includegraphics[width=7cm]{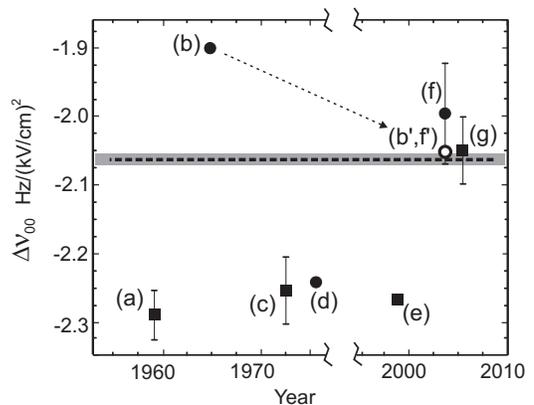}
\caption{The static Stark shift of the clock transition frequency.
The squares ({\tiny$\blacksquare$}) represent experimental values
of Haun et~al.~\cite{haun_zacharias}(a), Mowat~\cite{mowat}(c),
Simon et~al.~\cite{simon}(e), and Godone et~al.~\cite{godone}(g).
The dots (\textbullet) represent theoretical values of Feichtner
et~al.~\cite{feichtner}(b), Lee et~al.~\cite{lee}(d), and
Micalizio et~al.~\cite{micalizio}(f). The circle ($\circ$) (b',f')
represents the rescaled values of (b,f) as explained in the text.
The error bar of point (e) is smaller than the symbol size. The
dotted horizontal
line shows our result with its uncertainty (shaded band).}
\label{fig:scalpolstory}
\end{figure}

The first term in Eq.~(\ref{thirdordergeneral}) is not restricted
to diagonal matrix elements of the hyperfine interaction.
We have therefore extended the numerical evaluation of
Eq.~(\ref{thirdordergeneral}) by including \emph{off-diagonal}
terms.
Figure~\ref{fig:offdiagmatel} gives a schematic overview of all
possible configurations which include off-diagonal hyperfine matrix
elements and which are compatible with the hyperfine and Stark
operator selection rules.
It is interesting to note that the diagrams $\textbf{1}$ and
$\textbf{2}$ were already considered by Feichtner
et~al.~\cite{feichtner} in their calculation of the clock transition
Stark shift.
For unknown reasons, the off-diagonal terms were never included in the
calculation of the tensor polarizability.

We have evaluated all the diagrams shown in
Fig.~\ref{fig:offdiagmatel}, and in particular the diagrams
$\textbf{3}$, $\textbf{4}$, and $\textbf{5}$, which, to our
knowledge, were never considered before.
As noted in the figure, all diagrams lead to $F$ and $M$ dependent
energy shifts, except for diagram $\textbf{1}$, which gives an $M$
independent shift.
Thus all diagrams contribute, together with the diagonal
contributions $\textbf{A}$ and $\textbf{B}$ of
Fig.~\ref{fig:diagmatel}, to the Stark shift of the clock
transition, while only the diagrams $\textbf{B}$ and
$\textbf{2}\mbox{--} \textbf{5}$ contribute to the tensor
polarizability.
Moreover, the relative importance of the diagrams for the two
effects of interest is quite different.
In the case of the clock shift, we find that 90\% (95\%) of the
total contribution (n=$6\mbox{--} 18$) comes from the diagrams
$\textbf{A}$ and $\textbf{1}$ evaluated for $n=6,7$ ($n=6, 7, 8$).
In this case the contributing (electric dipole and hyperfine)
matrix elements are directly or indirectly given by experimentally
measured quantities.
The diagonal hyperfine matrix elements are proportional to the
measured hyperfine splittings, while the off-diagonal hyperfine
matrix elements between $S$ states of different principal quantum
numbers $n$ can be expressed in terms of the geometrical averages
of the hyperfine splittings of the coupled states, a relation
which holds at a level of $10^{-3}$~\cite{dzuba_sqrtformula}.
This gives us a high level of confidence in our value of the clock
shift rate.
The off-diagonal matrix elements of the diagrams
$\textbf{2}\mbox{--}\textbf{5}$ cannot be traced back to
experimental quantities. We have calculated these matrix elements
using wave functions obtained from the Schr\"odinger equation for
a Thomas-Fermi potential, with corrections including dipolar and
quadrupolar core polarization as well as spin-orbit interaction
with a relativistic correction factor following~\cite{norcross}.
We included $nS_J$, $nP_J$, and $nD_J$ states up to $n=18$ in the
perturbation sum and obtained
\begin{equation}
\Delta\nu_{00}/\mathbb{E}^{2}=-2.06(1)\:\Hz/(\kV/\cm)^{2}\:,
\label{clockshiftvalue}
\end{equation}
for the shift of the clock transition frequency, and
\begin{equation}
\alpha_{2}^{(3)}(F=I\pm J)=\mp3.72(25)\times 10^{-2}\:\Hz/(\kV/\cm)^{2}\:,
\label{tenspolfinalvalue}
\end{equation}
for the tensor polarizability.
The results are shown as dotted lines in Figs.~\ref{fig:tenspolstory}
and~\ref{fig:scalpolstory} together with previous theoretical and
experimental results.
The uncertainty of our calculated values is indicated by the shaded
bands.
The relative uncertainty of the clock shift is significantly
smaller than that of the tensor polarizability due to the use of
experimental values, with relatively small uncertainties.
For the contributions which we calculated explicitly from the
Schr\"odinger wave functions we use the accuracy
(4\%$\mbox{--}$8\%) with which these wave functions reproduce
experimental dipole matrix elements and hyperfine splittings to
estimate the precision of our results.
More details will be given in a forthcoming
publication~\cite{ulzega}.
We can claim that the present calculation of $\alpha_2^{(3)}$
yields a good agreement with all experimental data.

The situation with the Stark shift of the clock transition
frequency is less clear as there are disagreeing experimental
values.
While the experimental results~\cite{haun_zacharias},
\cite{mowat}, and~\cite{simon} [(a), (c), and (e) in
Fig.~\ref{fig:scalpolstory}] are supported by the theoretical
value of~\cite{lee}(d), our present result is in excellent
agreement with the recent experimental value of Godone
et~al.~\cite{godone}(g) and with the calculation of Micalizio
et~al.~\cite{micalizio}(f).
The shift was also calculated by Feichtner et~al.~\cite{feichtner} in
an approximation using hydrogenic wave functions, neglecting
spin-orbit interactions, and considering only the diagrams \textbf{A,
B, 1}, and \textbf{2} [point (b) in Fig.~\ref{fig:scalpolstory}].
With these approximations the scalar polarizability $\alpha_0$ can
be factored out of their final result.
We have rescaled point (b) in Fig.~\ref{fig:scalpolstory} using
more precise (consistent) values of $\alpha_0$ given
in~\cite{amini_scalpol,derevianko_scalpol,zhou_scalpol}, yielding
point (b') which is then consistent with the present result.
We also rescaled point (f) in the same manner, yielding (f') which
cannot be distinguished from (b').

We conclude by recalling the relevance of the latter result for
primary frequency standards.
One of the leading systematic shifts of the cesium clock frequency
is due to the interaction of the atoms with the blackbody
radiation (BBR) field.
It was shown~\cite{itano} that the dynamic BBR shift can be
parametrized in terms of the static Stark shift investigated here.
The correction factor for the BBR shift commonly used is based on
the precise experimental value of Simon et~al.~\cite{simon} [point
(e) in Fig.~\ref{fig:scalpolstory}], whose difference with the
present result is 21 (53) times the corresponding theoretical
(experimental) uncertainty.
In order to shine more light on this important issue we are preparing
an alternative new experiment for measuring the static Stark shift of
the clock transition frequency.

The authors acknowledge a partial funding of the present work by
Swiss National Science Foundation (grant 200020--103864).  One of us
(A.W.) thanks M.-A.~Bouchiat for useful discussions.

\bibliography{Ulzega_Stark_Bib}
\end{document}